\documentclass[fleqn]{annalen}
\usepackage{graphics}
\begin{document}
\newcommand{\volume}{8}              
\newcommand{\xyear}{1999}            
\newcommand{\issue}{5}               
\newcommand{\recdate}{29 July 1999}  
\newcommand{\revdate}{dd.mm.yyyy}    
\newcommand{\revnum}{0}              
\newcommand{\accdate}{dd.mm.yyyy}    
\newcommand{\coeditor}{ue}           
\newcommand{\firstpage}{000}         
\newcommand{\lastpage}{000}          
\setcounter{page}{\firstpage}        
\newcommand{\keywords}{quantum Hall effect, quantum spin chains, 
metal-insulator transitions} 
\newcommand{\PACS}{73.40.Hm, 71.30.+h, 75.10.Jm}
\newcommand{\shorttitle}{Shan-Wen Tsai et al., Study of critical behavior of 
the supersymmetric spin chain...} 
\title{Study of critical behavior of the supersymmetric spin chain that models plateau transitions in the integer quantum Hall effect}
\author{Shan-Wen Tsai and J.\ B.\ Marston} 
\newcommand{\address}
  {Department of Physics, Brown University,
  Providence, RI 02912-1843, USA}
\newcommand{\email}{\tt tsai@barus.physics.brown.edu} 
\maketitle
\begin{abstract}
We use the density-matrix renormalization-group (DMRG) 
algorithm and finite-size scaling to study a supersymmetric (SUSY) spin chain 
that models plateau transitions in the integer quantum Hall effect. To 
illustrate the method, we first present results for the $S=1$ 
antiferromagnetic spin chain. Different scaling behavior of the 
dimerization operator is obtained for the Haldane phase, the dimerized 
phase, and at the critical point separating the two phases. We then 
investigate the supersymmetric spin chain. As the on-site Hilbert space 
has infinite dimension, it is non-compact. We present DMRG calculations 
for supersymmetric truncations of the on-site Hilbert space and also for 
non-supersymmetric truncations. These two cases scale differently, in 
accord with a Lieb-Schultz-Mattis type theorem.
\end{abstract}
\section{Introduction}
\label{intro}
Quantum critical points are characterized by fluctuations over all length 
and time scales and by the appearance of power law scaling. Transitions 
between plateaus in the integer quantum Hall effect (IQHE) provide 
the clearest experimental example of quantum critical behavior in a 
disordered system. 

Critical behavior of the  supersymmetric (SUSY) spin chain that models 
plateau transitions in the IQHE is studied here using the 
density-matrix renormalization-group (DMRG) algorithm \cite{White} and 
finite-size scaling. In Sec.\ \ref{DMRG} we briefly describe the method 
\cite{Tsai}, show results for the rather well known $S=1$ antiferromagnetic 
spin chain, and then analyze the SUSY chain. In Sec.\ \ref{discussion} we 
discuss the results, which accord 
with the predictions of a generalized LSM theorem.
\section{DMRG results}
\label{DMRG} 
We use the so-called ``infinite-size'' DMRG algorithm \cite{White} as it is 
particularly simple to implement. It is a systematic way of building up 
a one-dimensional chain to calculate its low energy properties. At each step, 
two sites with on-site Hilbert space $D$ are added at the center of the chain. 
The left and right halves of the chain are considered as separate blocks, 
and the $M$ most probable states (as determined by the eigenvalues of the 
reduced density matrix) are retained for each block.
\subsection{$S=1$ chain}
\label{S=1}
Consider the general nearest-neighbor Hamiltonian for an isotropic $S=1$ 
quantum antiferromagnetic Heisenberg spin chain:
\begin{equation}
H = \sum_{j=0}^{L-2} \big{[} \cos \theta ~ ( \vec{S}_j \cdot \vec{S}_{j+1} ) +
\sin \theta ~ ( \vec{S}_j \cdot \vec{S}_{j+1} )^2 \big{]}.
\label{s1_hamilt}
\end{equation}
Dimerization, defined here as $\Delta(j) ~=~ \left| ~\langle 
S^x_j S^x_{j+1} + S^y_j S^y_{j+1}  \rangle -
 \langle S^x_{j-1} S^x_j + S^y_{j-1} S^y_j \rangle ~\right|$,
is induced in the interior of the chain by the open boundary conditions. 
We monitor the decrease of the induced dimerization at the center of the 
chain $\Delta(L/2)$ as the chain length $L$ is enlarged via the DMRG algorithm.
At a critical point, we expect power-law decay, with possible logarithmic 
and additive higher order corrections \cite{deGennes,Igloi}. In contrast, 
exponential decay or constant dimerization is expected for gapped systems. 

\begin{figure}[h]
\begin{tabular}[h]{lrr}
\resizebox{5.7cm}{5cm}{\includegraphics{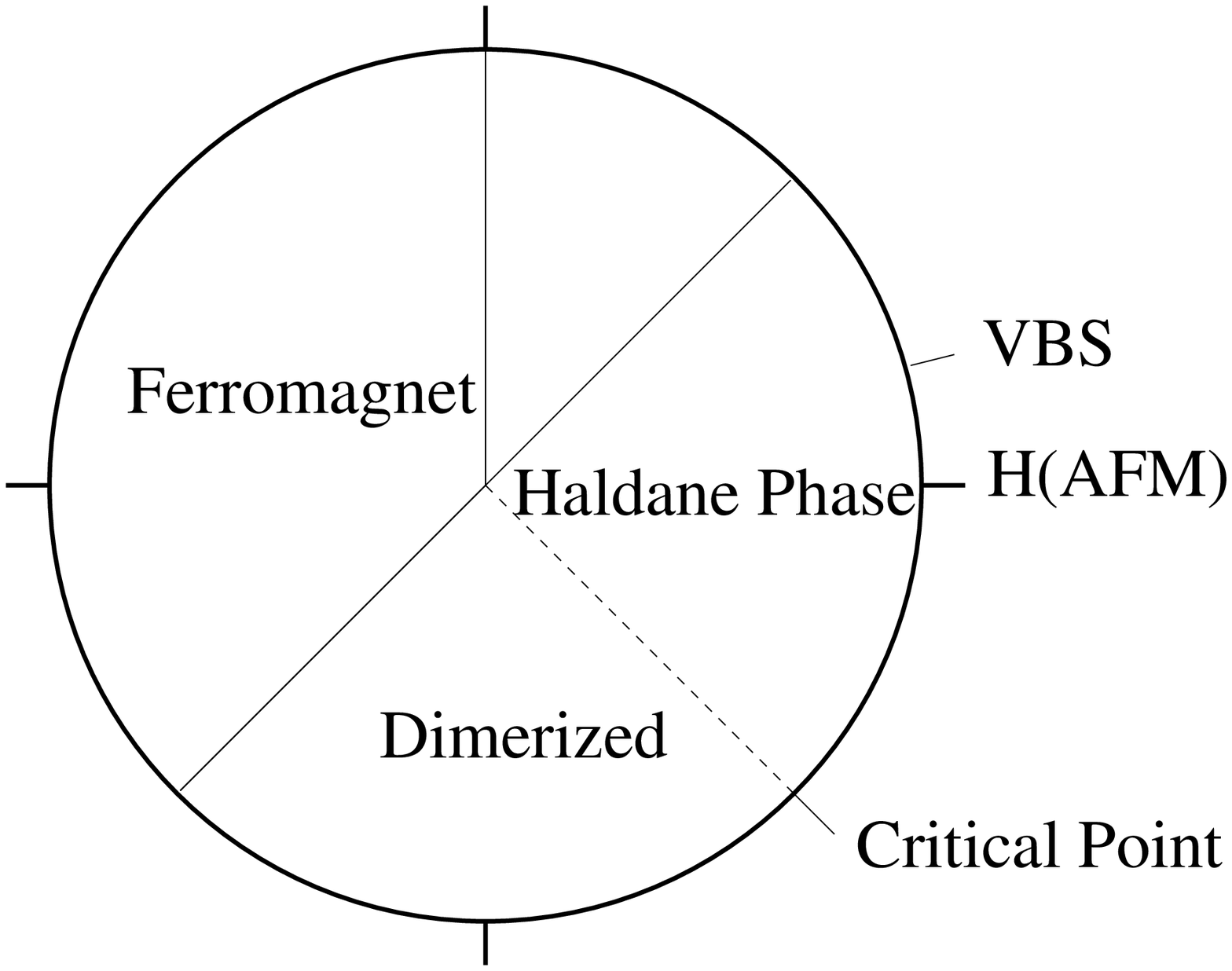}} &
\ \ \ \ \ \  &
\resizebox{6cm}{5cm}{\includegraphics{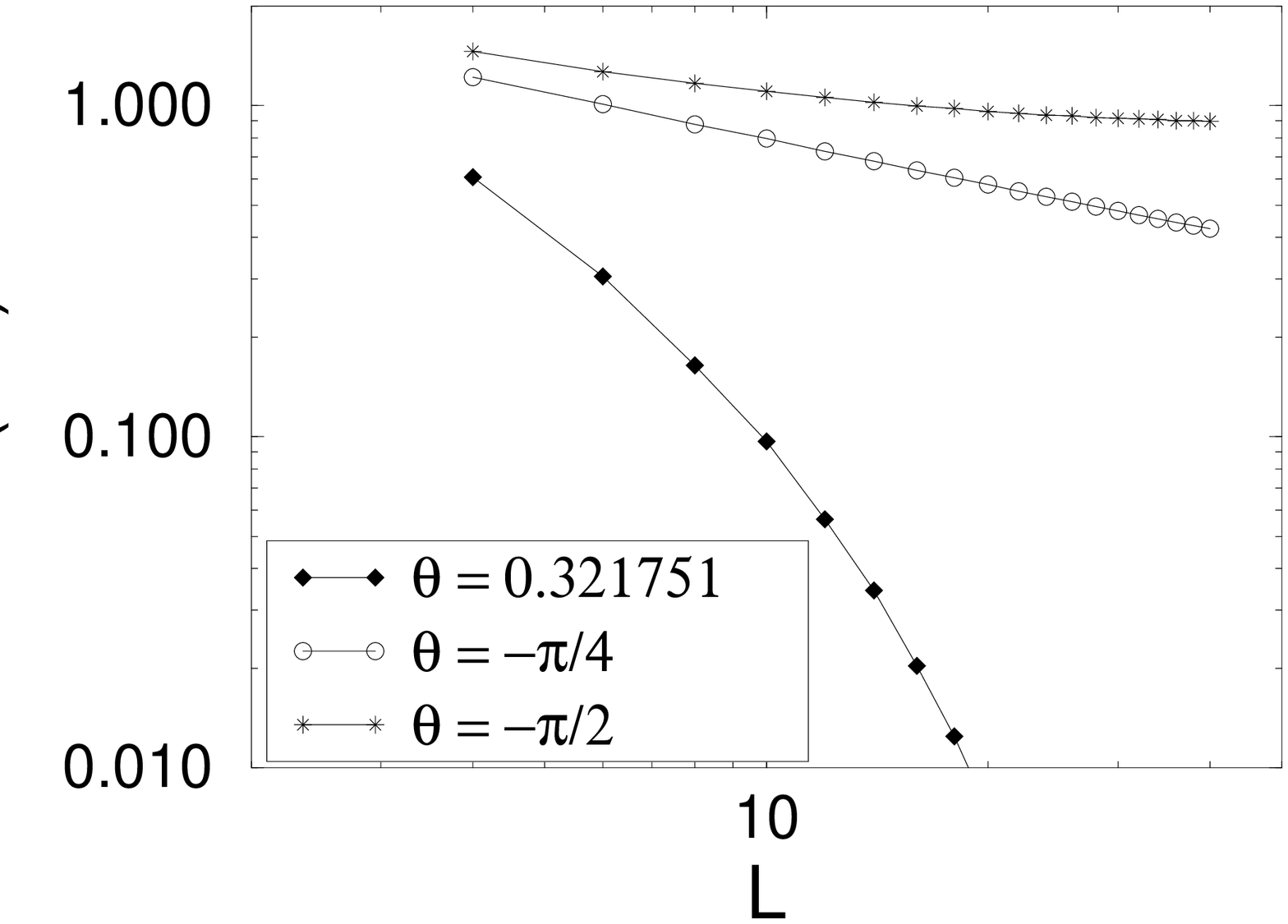}}
\end{tabular}
\caption{S=1 chain: (a) phase diagram and (b) induced dimerization at the 
center of the chain for $\theta$ in the Haldane phase, at the critical point 
and in the dimerized phase.}
\label{s1_fig}
\end{figure}

The phase diagram of the spin-1 chain (Eq.\ \ref{s1_hamilt}) can be 
represented on a circle where the parameter $\theta$ corresponds to the 
polar angle, as shown in Fig.\ \ref{s1_fig} (a).
Fig.\ \ref{s1_fig} (b) is a log-log plot of the induced dimerization at the 
center of the chain for three values of $\theta$. Power-law decay 
occurs at $\theta=-\pi/4$, the critical point. The point 
$\theta=\tan^{-1}(1/3)\approx0.321751$ 
corresponds to the exactly solvable Affleck-Kennedy-Lieb-Tasaki valence bond 
solid \cite{aklt} 
which is in the Haldane phase \cite{Haldane}. The system is gapped and the 
induced dimerization decays exponentially to zero. The $\theta=-\pi/2$ 
point is also gapped but in this case the induced dimerization decays to a 
constant since it is in the dimerized phase and translational symmetry is 
spontaneously broken (making $\sin \theta$ more negative favors the 
concentration of singlet correlations on isolated dimers). 
\subsection{Supersymmetric chain}
\label{susy}
Plateau transitions in the integer quantum Hall effect can be described by 
a quantum tunneling network model introduced by Chalker and 
Coddington \cite{Chalker}. In the anisotropic limit this model can be 
represented \cite{Dung-Hai,Ziqiang} by an independent-particle 
Hamiltonian which describes a chain of edge states alternating in 
propagation forward and backward in imaginary time. The method of 
supersymmetry (SUSY) \cite{McKane,Parisi,Efetov,Ian,Zirnbauer,
Balents} can then be used to perform the disorder average of the corresponding 
functional integral \cite{Zirnbauer2,Kondev}. Two spin species are introduced 
to permit the calculation of the disorder-averaged product of retarded and 
advanced Green's functions which determines the conductivity. The resulting 
effective Hamiltonian \cite{Marston} describes interacting spin-up and 
spin-down fermions and bosons, and can be written in terms of 16 spin 
operators. At the transition point between plateaus, a generalization of 
the theorem of Lieb, Schultz and Mattis shows that the system is quantum 
critical \cite{Marston}.

\begin{figure} [h]
\begin{tabular}[h]{lrr}
\resizebox{6cm}{5cm}{\includegraphics{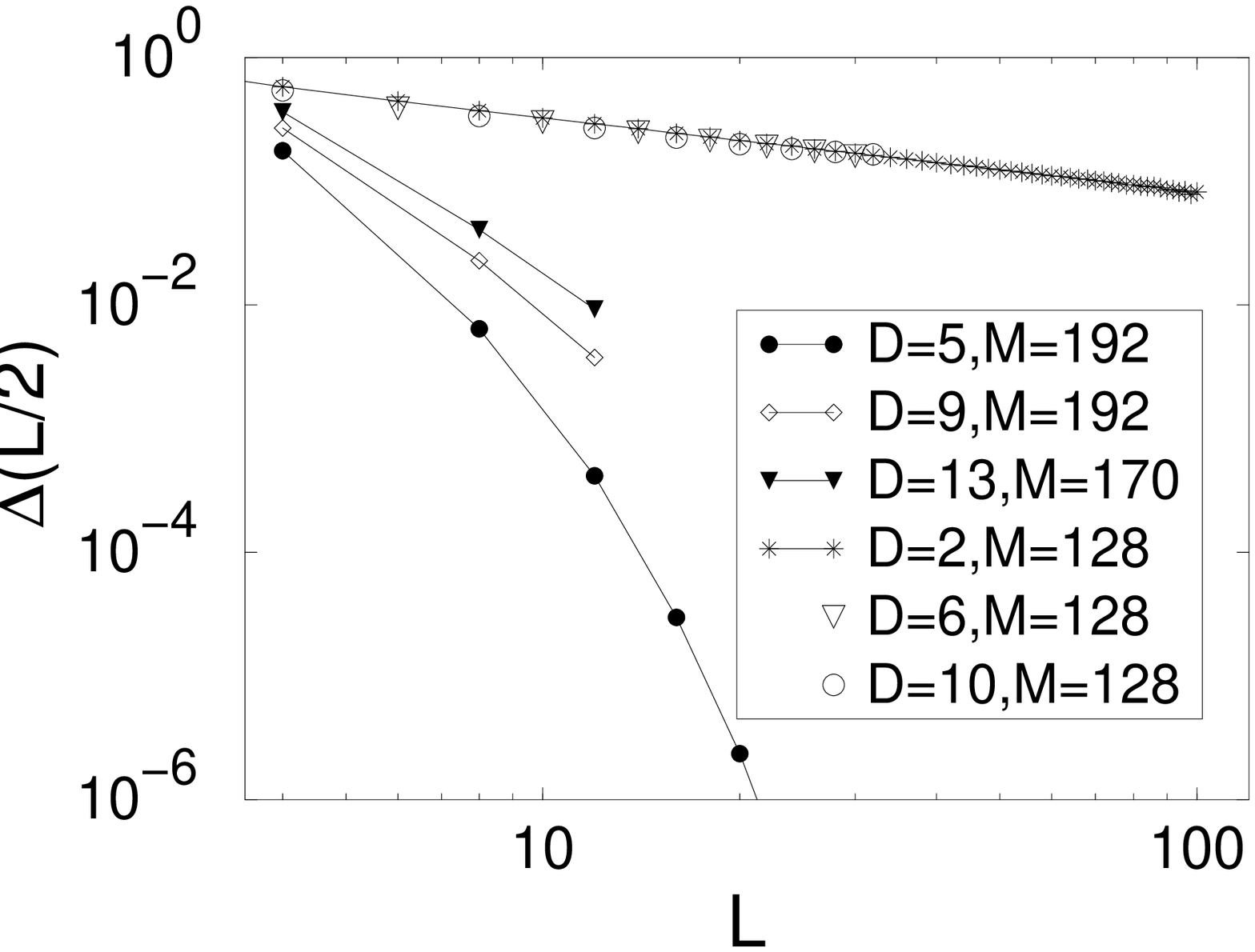}} & \ \ \ \  &
\resizebox{6cm}{5cm}{\includegraphics{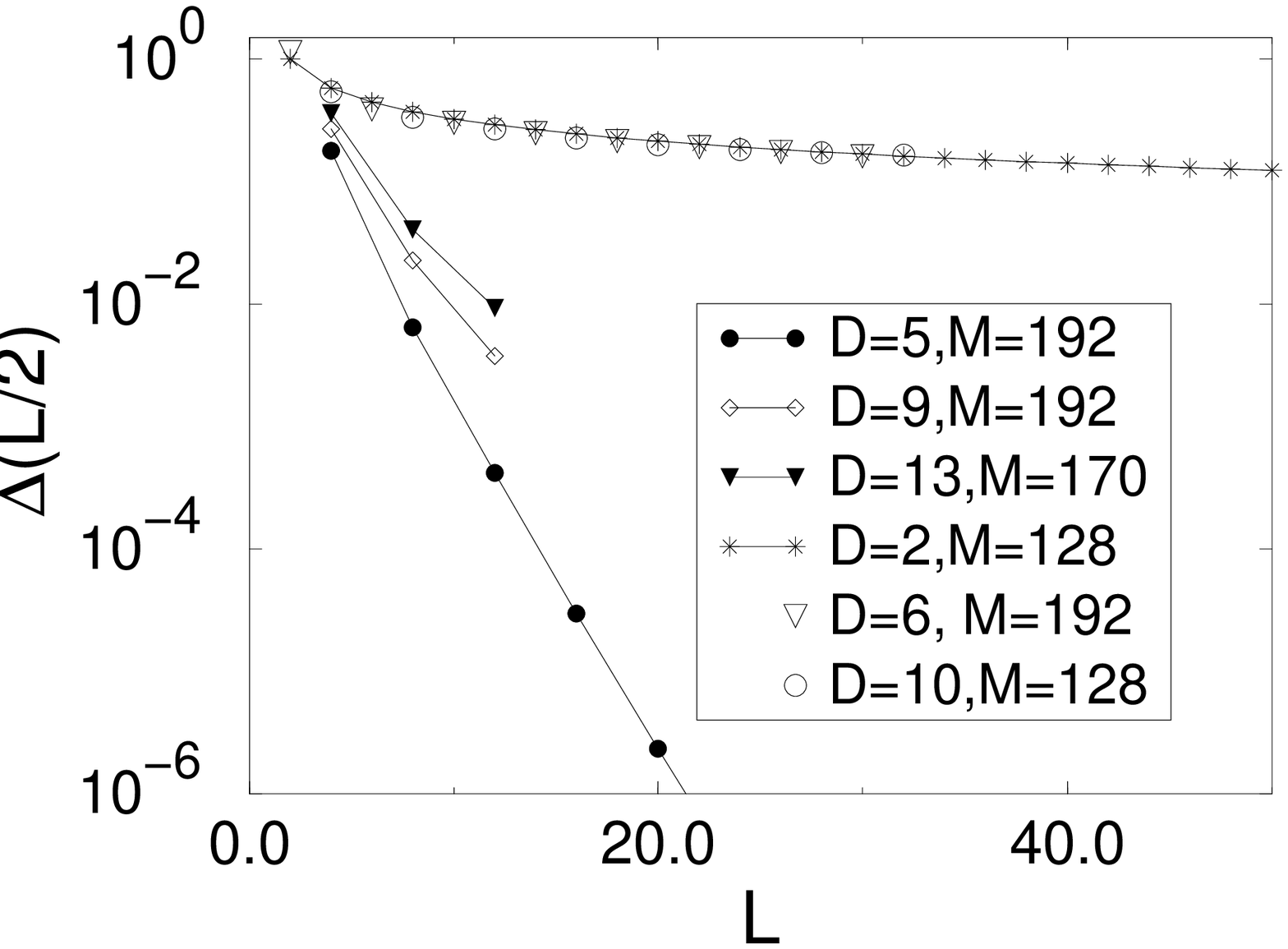}}
\end{tabular}
\caption{Log-log (a) and semi-log (b) plots of the induced dimerization at 
the center of the chain for SUSY truncations 
$D=5, 9, 13$ and for non-SUSY truncations $D=2, 6, 10$.}
\label{iqhe_fig}
\end{figure}

The SUSY spin chain that models plateau transitions in the IQHE 
is non-compact \cite{Zirnbauer2,Kondev} since an 
arbitrarily large number of bosons on each site is permitted. To employ the 
DMRG method, a truncation of the on-site Hilbert space is required. 
Introducing the integer level index $n = 0, 1, 2, \ldots$ we construct the 
on-site Hilbert space by adding to the vacuum state $|0 \rangle$ a tower of 
states built out of the quartets \cite{Marston}:
\begin{eqnarray}
\begin{array}{ll}
|4 n + 1 \rangle \equiv \frac{1}{n!}
(b^\dagger_\uparrow b^\dagger_\downarrow)^n
c^\dagger_\uparrow c^\dagger_\downarrow |0 \rangle \ , &
|4 n + 2 \rangle \equiv \frac{1}{\sqrt{n! (n+1)!}}
(b^\dagger_\uparrow b^\dagger_\downarrow)^n
b^\dagger_\uparrow c^\dagger_\downarrow |0 \rangle \ , \\
|4 n + 3 \rangle \equiv \frac{1}{\sqrt{n! (n+1)!}}
(b^\dagger_\uparrow b^\dagger_\downarrow)^n
c^\dagger_\uparrow b^\dagger_\downarrow |0 \rangle \ , &
|4 n + 4 \rangle \equiv \frac{1}{(n+1)!}
(b^\dagger_\uparrow b^\dagger_\downarrow)^n
b^\dagger_\uparrow b^\dagger_\downarrow |0 \rangle \ .
\end{array}
\label{Hilbert}
\end{eqnarray}
Truncations with $D=4n+1$ states contain full quartets and thus preserve 
SUSY. Also considered are 
non-SUSY truncations $D=4n+2$, with the state $|4n+1\rangle$ 
selected as the final state at the top of the tower. Fig.\ \ref{iqhe_fig} 
shows the induced dimerization at the center of the chain for SUSY truncations 
$D=5, 9, 13$ and non-SUSY truncations $D=2, 6, 10$. Calculations were done at 
the transition point (the dimerization parameter $R=0$, and the imaginary
frequency, which defines advanced and retarded propagators, is small, 
$\eta=10^{-4}$). As $R\rightarrow0$, the correlation length diverges 
as $\xi \sim R^{-\nu}$.
\section{Discussion of results}
\label{discussion}
In Ref. \cite{Marston}, we showed that the supersymmetric chain is quantum 
critical at $D\rightarrow\infty$ in the thermodynamic limit. Closely 
following the theorem of Lieb, Schultz and Mattis \cite{LSM} for 
half-odd-integer spin antiferromagnets, we considered a slow-twist 
operator $U$ which satisfies $U^{\dagger}[H,U]=O(1/L)$. Now $U$ is invariant 
under global supersymmetry rotations and therefore does not create 
low-energy excitations for truncations which respect SUSY. 
However, chains with non-SUSY truncations $D=4n+2$ are gapless in 
the thermodynamic limit. Continuity in the $D\rightarrow\infty$ limit 
requires that both truncations converge to the gapless behavior . This 
result was checked by DMRG calculations of the gap for different 
truncations and system sizes \cite{Marston}. The method used here, however, 
only requires the ground state as critical behavior can be extracted through 
the finite-size scaling analysis of the induced dimerization at the chain 
center.

DMRG results for the $S=1$ chain show power-law behavior at the critical 
point. Power-law behavior is also obtained for the superspin chain with 
non-SUSY truncations $D=4n+2$, as expected from the generalized LSM theorem. 
The special case $D=2$ corresponds to the ordinary $S=1/2$ Heisenberg 
antiferromagnet. Fig. \ref{iqhe_fig} (a) shows 
that non-SUSY truncations seem to be in the same universality class as 
the $S=1/2$ antiferromagnet. SUSY truncations are gapped and show 
exponential decay of the induced dimerization, as can be seen from the 
semi-log plot of Fig. \ref{iqhe_fig} (b). From the slopes in the semi-log 
plot, the relative sizes of the gap can be inferred for different 
SUSY $D=4n+1$ truncations. The gap decreases as the SUSY truncation size $D$ 
increases, and it must vanish in the physically relevant limit 
$D\rightarrow\infty$. 

\vspace*{0.25cm} \baselineskip=10pt{\small \noindent This research was 
supported in part by the NSF under Grants Nos. DMR-9357613, DMR-9712391 and 
PHY94-07194. Computations were carried out in C++ on Cray PVP machines at 
the Theoretical Physics Computing Facility at Brown University.}


\begin{thebibliography}{99}
\bibitem{White}S. R. White, Phys. Rev. Lett. {\bf 69} (1992) 2863;
Phys. Rev. B {\bf 48} (1993) 345

\bibitem{Tsai}S.-W. Tsai and J. B. Marston, unpublished

\bibitem{deGennes}M. E. Fisher and P.-G. de Gennes, C. R. Acad. Sc. Paris
B {\bf 287} (1978) 207

\bibitem{Igloi}F. Igloi and H. Rieger, Phys. Rev. Lett. {\bf 78} 
(1997) 2473

\bibitem{aklt}I. Affleck, T. Kennedy, E. H. Lieb and H. Tasaki, Phys. Rev.
Lett. {\bf 59} (1987) 799

\bibitem{Haldane}F. D. M. Haldane, Phys. Rev. Lett. {\bf 50} (1983) 1153;
Phys. Lett. {\bf 93A} (1983) 464

\bibitem{Chalker}J. T. Chalker and P. D. Coddington, J. Phys. C {\bf 21}
 (1988) 2665

\bibitem{Dung-Hai}D.-H. Lee, Phys. Rev. B {\bf 50} (1994) 10,788
 
\bibitem{Ziqiang}D.-H. Lee and Z. Wang, Phil. Mag. Lett. {\bf 73} (1996) 145

\bibitem{McKane}A. McKane, Phys. Lett. {\bf 76A} (1980) 22

\bibitem{Parisi}G. Parisi and N. Sourlas, J. Phys. (Paris) {\bf 41} 
(1980) L403

\bibitem{Efetov}K. B. Efetov, Adv. Phys. {\bf 32} (1983) 53

\bibitem{Ian}I. Affleck, J. Phys. C {\bf 17} (1984) 2323

\bibitem{Zirnbauer}M. R. Zirnbauer, Annalen der Physik {\bf 3} (1994) 513

\bibitem{Balents}L. Balents, M. P. A. Fisher, and M. Zirnbauer,
Nucl. Phys. {\bf B483} (1997) 601

\bibitem{Zirnbauer2}M. R. Zirnbauer, J. Math. Phys. {\bf 38} (1997) 2007

\bibitem{Kondev}J. Kondev and J. B. Marston, Nucl. Phys. {\bf B497} (1997) 639

\bibitem{Marston} J. B. Marston and S.-W. Tsai, Phys. Rev. Lett. {\bf 82} 
(1999) 4906
 
\bibitem{LSM} E. Lieb, T. D. Schultz, and D. C. Mattis, Ann. Phys. {\bf 16} 
(1961) 407

\end{thebibliography}
\end{document}